\documentclass[a4paper]{jpconf}
\usepackage{graphicx}
\usepackage{amssymb}
\begin{document}
\title{Cosmology with running parameters}

\author{Joan Sol\`a}

\address{HEP Group, Dep. ECM Universitat de Barcelona, and CER for Astrophysics,
Particle Physics and Cosmology, Av. Diagonal 647,  08028
Barcelona, Catalonia, Spain}

\ead{sola@ifae.es}

\begin{abstract}
The experimental evidence that the equation of state (EOS) of the
dark energy (DE) could be evolving with time/redshift (including
the possibility that it might behave phantom-like near our time)
suggests that there might be dynamical DE fields that could
explain this behavior. We propose, instead, that a variable
cosmological term (including perhaps a variable Newton's
gravitational coupling too) may account in a natural way for all
these features.
\end{abstract}

\section{Introduction}

The accelerated expansion of the universe is nowadays one of the
central issues of observational and theoretical cosmology. The
usual paradigm assumes the existence of the so-called {\em dark
energy} (DE)-- a mysterious cosmic component with negative
pressure. An obvious candidate for the role of DE is the
cosmological constant (CC) \cite{CCP}. Others are the dynamical DE
models\,\cite{PSW}, e.g. quintessence \cite{Quintessence}, phantom
energy \cite{phantom} etc.  Here we generalize the cosmological
constant concept allowing the CC term $\rho_{\Lambda}$ (with
dimensions of energy density) and possibly the gravitational
coupling ($G$) being variable with the cosmic time, both
assumptions compatible with the cosmological principle. The
support for such a generalization comes from quantum field theory
on the curved space-time
\cite{JHEPCC1,RGTypeIa,ShapSolNPB,RGTypeIa2,Babic,SSS} and/or
quantum gravity approaches \cite{Reuter}.

\section{Two cosmological pictures: standard DE versus variable cosmological term}
The usual dark energy picture assumes the existence of two
separately conserved cosmological components, the
matter-radiation component and the DE component. Not necessarily
so in the variable $\rho_{\Lambda}/G$ picture (for short, CC
picture). The general Bianchi identity of the Einstein tensor
leads to $\label{eq:covBianchi} \nabla^{\mu} \left[G (T_{\mu \nu}
+ g_{\mu \nu} \rho_{\Lambda}) \right]=0$, which for FRW metric
implies
\begin{equation}\label{BianchiGeneral}
\frac{d}{dt}\,\left[G(\rho+\rho_{\Lambda})\right]+3\,G\,H\,(\rho+p)=0\,.
\end{equation}
This ``mixed" conservation law connects the variation of
$\rho_{\Lambda}$, $G$ and $\rho$, and hence the evolution of the
matter density $\rho$ may be noncanonical.  A number of variable
CC models of various kinds \cite{CCvariable1}, and the
renormalization group (RG) models of running $\rho_{\Lambda}$ and
$G$ \cite{JHEPCC1,RGTypeIa,Babic,SSS,Reuter} provide the basis
for the variability of these cosmological parameters. The general
expression for the Hubble parameter in the CC picture in the
matter ($\alpha=3$) or radiation ($\alpha=4$) epochs is
\begin{equation}\label{HLambda}
H^2_{CC}(z)=H^2_0\,\left[\Omega_{M}^0\,f_{M}(z)(1+z)^{ \alpha}
+\Omega_{\Lambda}^0\,f_{\Lambda}(z)\right]\,.
\end{equation}
Here $f_{M}$ and $f_{\Lambda}$ are certain functions  of redshift
(see e.g. the RG cosmological models\,\cite{SSS,SS1}), with
$f_{M}(0)=1$ and $f_{\Lambda}(0)=1$, in accordance with the cosmic
sum rule $\Omega_{M}^{0}+\Omega_{\Lambda}^{0}=1$.

\section{Matching of pictures and effective dark energy equation of state}

The two pictures presented in the preceding section are assumed to
be equivalent descriptions of the same cosmological evolution.
Their matching requires that the expansion history of the universe
is the same in both pictures, i.e. that their Hubble functions are
equal: $H_{DE}=H_{CC}$. The general Bianchi identity
(\ref{BianchiGeneral}) then leads to an effective EOS parameter
$\omega_e=p_{D}/\rho_{D}$ for the CC picture, given by
\begin{equation}\label{we2}
\omega_e(z)=-1+\frac{\alpha}{3}\,\left(1-\frac{\xi_{\Lambda}(z)}{\rho_D(z)}\right)\,,
\end{equation}
where $\xi_{\Lambda}(z)\equiv(G(z)/G_0)\,\rho_{\Lambda}(z)$.  The
effective DE density in the CC picture can be cast as
\begin{equation}\label{IF2}
\rho_D(z)=\xi_{\Lambda}(z)-\left(1+z\right)^{\alpha}\,
\int_{z^{*}}^z\frac{dz'}{(1+z')^{\alpha}}\frac{d\xi_{\Lambda}(z')}{dz'}\,.
\end{equation}
Here $z^*$ is a redshift where $\xi_{\Lambda}(z^*)=\rho_D(z^*)$.
From (\ref{we2}) it is clear that $\omega_e$ crosses the
$\omega_e = -1$ ``barrier'' just at $z=z^*$. Quite remarkably, one
can show that a value $z^*$ {\em always} exists near our present
time: namely in the recent past, immediate future or just at
$z^*=0$. The proof of this claim is obtained by straightforward
calculation starting from the matching condition and the
conditions that the general Bianchi identity
(\ref{BianchiGeneral}) imposes on functions $f_{\Lambda}$ and
$f_{M}$ in (\ref{HLambda})\,\cite{SS1}.
 Let us compute the  slope of the $\rho_{D}$ function (\ref{IF2}):
\begin{equation}\label{dIF2}
\frac{d\rho_D(z)}{dz}=-\alpha\,\left(1+z\right)^{\alpha-1}
\int_{z^{*}}^z\frac{dz'}{(1+z')^{\alpha}}\frac{d\xi_{\Lambda}(z')}{dz'}\,.
\end{equation}
This compact expression reveals some counterintuitive and general
aspects of the effective DE density evolution for a variable CC
model in which $\xi_{\Lambda}(z)$ is a monotonous function of
$z$. Thus, for $\xi_{\Lambda}(z)$ decreasing with $z$, $\rho_{D}$
behaves like quintessence for $z>z^*$, whereas for $z<z^*$ it
behaves phantom-like! (For a concrete framework, see Section
\ref{RGmodel} and Fig.\,\ref{plot}.)  Especially interesting
results are obtained when in the variable CC model the matter
component $\rho$ is separately conserved. In this case we have
${d\xi_{\Lambda}}/{dt}=-(\rho/G_0)\,{dG}/{dt}$, which results in
the following expression for the slope of $\rho_{D}$:
\begin{equation}\label{varG}
\frac{d\rho_D}{dz}=\alpha (1+z)^{\alpha-1} \,\frac{\rho(0)}{G_0}\
[G(z)-G(z^{*})]\,.
\end{equation}
Thus in this case the properties of $\rho_{D}$ depend only on the
scaling of $G$ with redshift, e.g. if $G$ is asymptotically free
and $z^{*}>z$, then $\rho_D$ behaves effectively as quintessence.

\section{Effective dark energy picture of a running $G$ and $\rho_{\Lambda}$ model}
\label{RGmodel}

\begin{figure}[t]
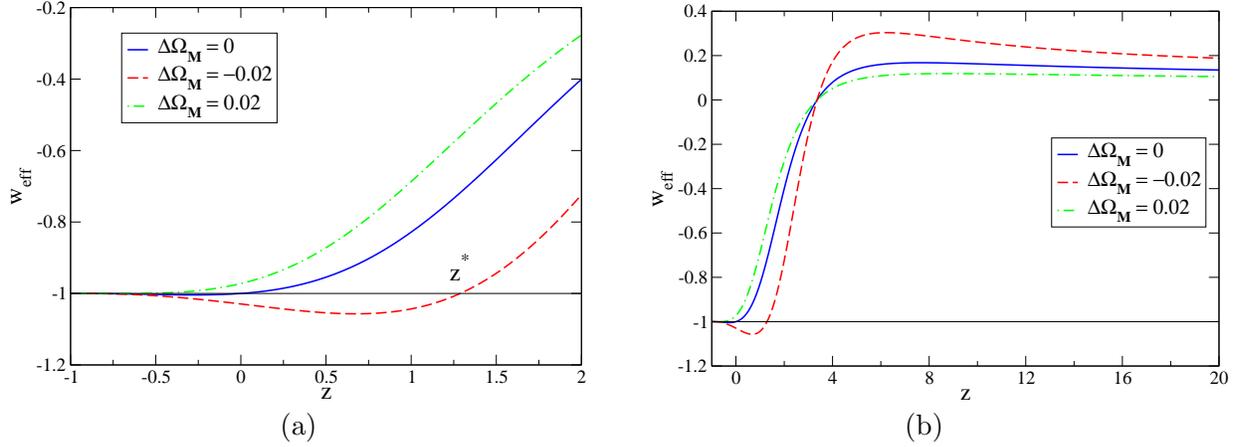

    \begin{tabular}{cc}
      \resizebox{0.48\textwidth}{!}{\includegraphics{fig1.eps}} &
      \hspace{0.3cm}
      \resizebox{0.48\textwidth}{!}{\includegraphics{fig2.eps}} \\
      (a) & (b)
    \end{tabular}
    \caption{\textbf{(a)} $\omega_e$,
Eq.\,(\protect\ref{wpflat1}), as a function of $z$ for various
$\Delta\Omega_M$ at fixed $\nu=-\nu_0$,
$\Omega_M^0=0.3\,,\Omega_{\Lambda}^0=0.7$; \textbf{(b)} Extended
$z$ range of the plot (a). $z^{*}$ is the crossing point of the
barrier $w=-1$.}
  \label{plot}
\end{figure}

As an illustration of the aforementioned procedure for obtaining
the effective dark energy properties of a variable CC model, in
this section we present the analysis of the renormalization group
model of \cite{RGTypeIa} characterized by $G =\mathrm{const}$ and
the CC evolution law $\rho_{\Lambda}=C_1+C_2 H^2$. Here
$C_1=\rho_{\Lambda,0}-(3 \nu H_0^2)/(8 \pi G)$ and $C_2=(3
\nu)/(8 \pi G)$, where $\nu$ is the single free parameter of the
model --a typical value is $|\nu|=\nu_0\equiv
1/12\pi$\,\cite{SS1}. In this particular model
$\xi_{\Lambda}(z)=\rho_{\Lambda}(z)$. Therefore for the flat
universe case the effective parameter of EOS of this running
$\rho_{\Lambda}(z)$ model one finds
\begin{equation}\label{wpflat1}
\hspace{-2cm} \omega_e(z)\left|_{\Delta\Omega\neq 0}\right.
=-1+(1-\nu)\,\frac{\Omega_M^0\,(1+z)^{3(1-\nu)}-\tilde{\Omega}_M^0\,(1+z)^3}
{\Omega_M^0\,[(1+z)^{3(1-\nu)}-1]-(1-\nu)\,[\tilde{\Omega}_M^0\,(1+z)^3-1]}\,.
\end{equation}
Here $\Delta\Omega_{M}\equiv
\Omega_{M}^0-\tilde{\Omega}_{M}^0\neq 0$ is the difference of
parameters in the two pictures, corresponding to two different
fits of the same data. For $|\nu|\ll 1$ we may expand the previous
result in first order in $\nu$. Assuming (conservatively) that
$\Delta\Omega_M=0$ we find
\begin{equation}\label{expwq}
\omega_e(z)\simeq-1-3\,\nu\frac{\Omega_M^0}{\Omega_{\Lambda}^0}\,(1+z)^3\,\ln(1+z)\,.
\end{equation}
This result reflects the essential qualitative features of the
analysis presented in the previous section, where in this case
$z^{*}=0$. For $\nu>0$, Eq.\,(\ref{expwq}) shows that we can get
an (effective) phantom-like behavior ($\omega_e<-1$), and for
$\nu<0$ we can have (effective) quintessence behavior. We see
that this variable CC model can give rise to two types of very
different behaviors by just changing the sign of a single
parameter. In Fig.\ref{plot} we show a more general case where
$z^{*}\gtrsim 0$, corresponding to $\Delta\Omega_M\neq 0$. In
contrast to the previous situation, here the variable CC model may
exhibit phantom behavior for $\nu<0$ (if $\Delta\Omega_M<0$), and
manifests itself through the existence of a transition point
$z^{*}$ in our recent past  -- marked explicitly in the figure.

\section{Conclusions}
We have shown that a model with variable $\rho_{\Lambda}$ (and may
be also with variable $G$) generally leads to a non-trivial
effective EOS; the model can effectively appear as quintessence,
and even as phantom energy, without need of invoking any
combination of fundamental quintessence and phantom fields. This
possibility should be taken into account in the next generation of
high precision cosmology experiments aiming to determine the
effective EOS of the DE.

\ack This work has been supported in part by MEC, FEDER and DURSI.
It is my pleasure to thank H. \v{S}tefan\v{c}i\'{c} for a very
nice collaboration and discussion of these matters.

\end{document}